\documentclass[%
 aip,
 amsmath,amssymb,
 reprint,%
]{revtex4-1}
\usepackage{url}
\usepackage{graphicx}
\usepackage{dcolumn}
\usepackage{bm}

\usepackage[utf8]{inputenc}
\usepackage[T1]{fontenc}
\usepackage{mathptmx}
\usepackage{etoolbox}
\usepackage[dvipsnames]{xcolor}

\makeatletter
\def\@email#1#2{%
 \endgroup
 \patchcmd{\titleblock@produce}
  {\frontmatter@RRAPformat}
  {\frontmatter@RRAPformat{\produce@RRAP{*#1\href{mailto:#2}{#2}}}\frontmatter@RRAPformat}
  {}{}
}%
\makeatother
\RequirePackage[normalem]{ulem} 
\RequirePackage{color}\definecolor{RED}{rgb}{1,0,0}\definecolor{BLUE}{rgb}{0,0,1} 
\providecommand{\DIFadd}[1]{{\protect\color{blue}\uwave{#1}}} 
\providecommand{\DIFaddbegin}{} 
\providecommand{\DIFaddend}{} 
\providecommand{\DIFdelbegin}{} 
\providecommand{\DIFdelend}{} 
\providecommand{\DIFaddbeginFL}{} 
\providecommand{\DIFaddendFL}{} 
\providecommand{\DIFdelbeginFL}{} 
\providecommand{\DIFdelendFL}{} 
\newcommand{\DIFscaledelfig}{0.5}
\RequirePackage{settobox} 
\RequirePackage{letltxmacro} 
\newsavebox{\DIFdelgraphicsbox} 
\newlength{\DIFdelgraphicswidth} 
\newlength{\DIFdelgraphicsheight} 
\LetLtxMacro{\DIFOincludegraphics}{\includegraphics} 
\newcommand{\DIFaddincludegraphics}[2][]{{\color{blue}\fbox{\DIFOincludegraphics[#1]{#2}}}} 
\newcommand{\DIFdelincludegraphics}[2][]{
\sbox{\DIFdelgraphicsbox}{\DIFOincludegraphics[#1]{#2}}
\settoboxwidth{\DIFdelgraphicswidth}{\DIFdelgraphicsbox} 
\settoboxtotalheight{\DIFdelgraphicsheight}{\DIFdelgraphicsbox} 
\scalebox{\DIFscaledelfig}{
\parbox[b]{\DIFdelgraphicswidth}{\usebox{\DIFdelgraphicsbox}\\[-\baselineskip] \rule{\DIFdelgraphicswidth}{0em}}\llap{\resizebox{\DIFdelgraphicswidth}{\DIFdelgraphicsheight}{
\setlength{\unitlength}{\DIFdelgraphicswidth}
\begin{picture}(1,1)
\thicklines\linethickness{2pt} 
{\color[rgb]{1,0,0}\put(0,0){\framebox(1,1){}}}
{\color[rgb]{1,0,0}\put(0,0){\line( 1,1){1}}}
{\color[rgb]{1,0,0}\put(0,1){\line(1,-1){1}}}
\end{picture}
}\hspace*{3pt}}} 
} 
\LetLtxMacro{\DIFOaddbegin}{\DIFaddbegin} 
\LetLtxMacro{\DIFOaddend}{\DIFaddend} 
\LetLtxMacro{\DIFOdelbegin}{\DIFdelbegin} 
\LetLtxMacro{\DIFOdelend}{\DIFdelend} 
\DeclareRobustCommand{\DIFaddbegin}{\DIFOaddbegin \let\includegraphics\DIFaddincludegraphics} 
\DeclareRobustCommand{\DIFaddend}{\DIFOaddend \let\includegraphics\DIFOincludegraphics} 
\DeclareRobustCommand{\DIFdelbegin}{\DIFOdelbegin \let\includegraphics\DIFdelincludegraphics} 
\DeclareRobustCommand{\DIFdelend}{\DIFOaddend \let\includegraphics\DIFOincludegraphics} 
\LetLtxMacro{\DIFOaddbeginFL}{\DIFaddbeginFL} 
\LetLtxMacro{\DIFOaddendFL}{\DIFaddendFL} 
\LetLtxMacro{\DIFOdelbeginFL}{\DIFdelbeginFL} 
\LetLtxMacro{\DIFOdelendFL}{\DIFdelendFL} 
\DeclareRobustCommand{\DIFaddbeginFL}{\DIFOaddbeginFL \let\includegraphics\DIFaddincludegraphics} 
\DeclareRobustCommand{\DIFaddendFL}{\DIFOaddendFL \let\includegraphics\DIFOincludegraphics} 
\DeclareRobustCommand{\DIFdelbeginFL}{\DIFOdelbeginFL \let\includegraphics\DIFdelincludegraphics} 
\DeclareRobustCommand{\DIFdelendFL}{\DIFOaddendFL \let\includegraphics\DIFOincludegraphics} 
\RequirePackage{listings} 
\RequirePackage{color} 
\lstdefinelanguage{DIFcode}{ 
  moredelim=[il][\color{red}\sout]{\%DIF\ <\ }, 
  moredelim=[il][\color{blue}\uwave]{\%DIF\ >\ } 
} 
\lstdefinestyle{DIFverbatimstyle}{ 
	language=DIFcode, 
	basicstyle=\ttfamily, 
	columns=fullflexible, 
	keepspaces=true 
} 
\lstnewenvironment{DIFverbatim}{\lstset{style=DIFverbatimstyle}}{} 
\lstnewenvironment{DIFverbatim*}{\lstset{style=DIFverbatimstyle,showspaces=true}}{} 

\begin{document}

\title{Quantum State-Channel Duality for the calculation of Standard Model scattering amplitudes.\\}

\author{Clelia Altomonte,}
\affiliation{Department of Physics, Strand, King's College London,
WC2R 2LS}%
\author{Alan J. Barr}%
 \email{clelia.altomonte@kcl.ac.uk}
\affiliation{Department of Physics, Keble Road, University of Oxford, OX1 3RH\\
Merton College, Merton Street, Oxford, OX1 4JD}%

\date{21 September 2023, First submitted: 6 October 2022 (arXiv: 4530106)\\
qd   Accepted for publication in Physics Letters B: 30 October 2023}

\begin{abstract}
Abstract: Recent instances of successful application of quantum information techniques to particle physics problems invite for an analysis of the mathematical details behind such connection. In this paper, we identify the Choi-Jamiolkowski isomorphism, or state-channel duality, as a theoretical principle enabling the application of the theory of quantum information to the scattering  amplitudes associated with Standard Model processes.

\end{abstract}

\maketitle

\section{\label{sec:level1} QUANTUM INFORMATION THEORY APPLIED TO HIGH ENERGY SYSTEMS}

While the first suggestion to make simulations of Nature quantum mechanical famously dates back to Richard Feynman\cite{feynman_1982}, recent attempts to apply the theory of quantum information to the study of high energy physics systems have proven particularly successful. As a paradigmatic example quantum state tomography, a procedure that allows full reconstruction of the density matrix of a system by performing a complementary series of measurements on an ensemble of identical copies of the system under scrutiny \cite{nielsen2010quantum}, is ideally applicable to colliders, where large numbers of events are generated\cite{barr2022belltype,Fabbrichesi_2021,Afik_2021,ashbypickering2022quantum}, and has been applied to numerical simulation studies of various high energy particle physics systems\cite{Fabbrichesi_2021,Afik_2021,Barr_2022,ashbypickering2022quantum}. 
Quantum algorithms, including quantum machine learning techniques, have been developed for the recognition of Standard Model and beyond signatures in data\cite{M_tt_nen_2004,https://doi.org/10.48550/arxiv.2203.03578,Wu_2021}, and for a more computationally economic simulation of collider events\cite{gustafson2022collider}.

While these results verify the expected agreement between the two fields of particle physics and quantum information (being that the Standard Model is based on quantum field theory, which is a quantum theory), the mathematical details behind such connection could be further exploited, leading to novel insights into both fields. In this paper, we identify the Choi-Jamiolkowski isomorphism \cite{PhysRevA.87.022310}, or state-channel duality, as a theoretical principle enabling a systematic application of the theory of quantum information to the calculation of Standard Model scattering amplitudes, and consider it worthy of the attention of the particle physics community for the following reasons.\\

On the one hand, the utility of this approach lies in pointing to a mathematical relation in the form of an isomorphism to establish a rigorous dictionary connecting Standard Model scattering amplitudes and the outputs of quantum channels. The existence of such dictionary allows one to better exploit techniques from quantum information theory in particle physics processes which could be experimentally observed at working and near-term colliders, hence providing a mathematical framework that has the potential to address the following problems:

\noindent (i) testing the foundations of quantum mechanics in particle physics colliders, by identifying the collider itself as a natural quantum-information processing machine provided that quantum mechanics holds. In this paper, we address this problem by providing, as an example, a tree-level process which shows how the presented framework allows us to test this assumption;\\

\noindent(ii) performing tests of basic quantum predictions such as entanglement and quantum state and process tomography in high energy regimes (cf. refs.\mbox{
\cite{barr2022belltype,Barr_2022,ashbypickering2022quantum}}\hskip0pt
); \\

 \noindent(iii) expanding theoretical modelling into regions where the Standard Model theory is difficult to calculate perturbatively, and where phenomenological models are currently employed, by having an alternative method for performing calculations.
\\


On the other hand, the scope of our approach does not overlap with other quantum computing protocols currently applied to nuclear and particle physics, such as quantum Hamiltonian evolution simulations\mbox{
\cite{Klco_2022, Bauer:2023qgm, PRXQuantum.4.027001,Ba_uls_2020}}\hskip0pt
, where the full dynamics of the particle physics system is simulated with the aim of observing non-perturbative effects. Resorting only to the knowledge of initial and final state density matrices, our approach allows us to establish whether a quantum channel which is equivalent to a given S-matrix can be constructed. This makes it possible reliably to use quantum information tools to learn about the Hamiltonian underlying the channel, and can potentially reveal evidence of new physics beyond the Standard Model.\\
\\

 The present paper is organised as follows: in Section II, we assess the conditions for Standard Model scattering processes to be suitable systems for the theoretical and experimental study of quantum information, satisfying the general conditions for successful quantum computing implementation\cite{nielsen2010quantum}. In Section III, by applying state-channel duality we show that Standard Model scattering amplitudes can be calculated as quantum channels. In Section IV we present an example model to illustrate how helicity matrix elements associated with tree-level electroweak interaction Feynman diagrams can be reproduced by means of simple quantum channels. In Section V, we introduce a dictionary connecting some selected particle physics processes and quantum channels. Section VI is dedicated to the extension of state-channel duality to higher-order corrections. Finally, in Section VII we comment on some possible applications of the present work.

\section{\label{sec:level1}CONSTRUCTING A QUANTUM INFORMATION DESCRIPTION OF STANDARD MODEL SCATTERING AMPLITUDES}

A general formulation\cite{nielsen2010quantum} of the conditions for a system to successfully implement quantum computation requires that it must be possible to: \\
\textit{\\a. Robustly represent quantum information;}\\
\textit{\\b. Perform a universal family of unitary\\ transformations;}\\
\textit{\\c. Prepare a fiducial initial state;}\\
\textit{\\d. Measure the output result;}\\

 \noindent where `successful implementation' means that the system can be manipulated in such a way as to correspond to a given quantum circuit. It is also worth clarifying that particle physics scattering processes described by quantum field theory and observed at colliders can be considered to implement quantum computation, albeit in a passive way. In fact, while candidate systems for the construction of physical quantum computers (e.g. optical systems\cite{nielsen2010quantum}) should in general allow for an easy and more extensive manipulation of the system's parameters, the experimenters' possibility of intervention at colliders is limited to the possible use of polarised beams and in determining other limited experimental settings. However, constructing a rigorous quantum information description of Standard Model scattering amplitudes is possible because the four conditions outlined above are fulfilled. We consider each condition in the subsections below.

\subsection{Robustly representing quantum information: the choice of basis}

\textit{Condition a.} rests on being able to encode particle physics observables such as spin states into \textit{qudits} (i.e. quantum mechanical $d$-dimensional bits of information\cite{nielsen2010quantum}). This is verified by considering that in quantum information theory it is commonplace to represent the state of a qudit, a $d$-dimensional system described by a $d\times d$ density matrix, in vectorial form (Bloch vector).
Since the space of matrices is a vector space, it is possible to find bases matrices to decompose any matrix\cite{Bertlmann_2008}. 

As an example, in the generalized Gell-Mann matrix basis, the matrices correspond to the standard $SU(N)$ generators ($d = N$), and induces a Bloch vector with real components, which can readily be interpreted as expectation values of measurable quantities (even though the SU($d$) generators do not necessarily correspond to any physical state, combining them with the Identity operator can result in a linear combination which can represent a physical state density operator). Considering a $2$-dimensional state as an example, the Pauli operators offer a complete Hermitian operator basis to parametrise the one qubit space, and correspond to the Pauli group used to describe physical observables of spin-$\frac{1}{2}$ particles \cite{Dogra_2018,weyl_2014,PhysRev.125.1067}. Because particle physics systems' observables such as spin can also be described according to some group-theoretic structure, it follows that they can easily be connected to quantum state's Bloch vector  `robustly', making them very interesting qudit implementations. \\

It is worth noting that while spin is a particle physics observable that can in principle be regarded as a good candidate for encoding quantum information, having a both a physical interpretation and unitary representation, alternative basis choices may prove more useful when trying to connect the processes observed at colliders with a quantum information description (as an example, cf. Appendix A for a detailed explanation of the helicity basis used in Section IV).

\subsection{Performing a universal family of unitary transformations}

Because the Standard Model of particle physics is a quantum field theory and is constructed on the unitary product group $\mathrm{SU}(3) \times \mathrm{SU}(2) \times \mathrm{U}(1)$, and qubit quantum computers generally implement unitary operations, it is expected that an interpretation of the Standard model in terms of quantum information is achievable (thus satisfying \textit{ Condition b.}).

As we will show in the following section, state-channel duality can be used to prove a related point: even at the lowest perturbation order such translation is achievable and hence quantum information machinery and tests (e.g. Bell inequalities violation tests) can be rigorously applied to particle physics processes and give meaningful results.

\subsection{Preparing a fiducial initial state}

\textit{ Condition c.} is satisfied by the proposal \cite{Afik_2021} that accelerators, especially those that allow for spin polarised beams, can be used to study quantum information processes. On the quantum information side, as mentioned above, the choice of basis needs to be made carefully, eigenstates of unitary operators corresponding to physical observables are the preferred choice.

\subsection{Measuring the output result}

Finally, colliders fulfill \textit{ Condition d.} by construction, being designed and engineered for the purpose of precise and reliable energy and momentum measurements. We can take the past history of successful measurements carried out at colliders as proof of this. In addition, as explained in section IV below, specific particle physics processes (e.g. $t \bar{t}$ quarks leptonic decays) provide particularly good candidates for quantum information applications, allowing for reliable access to spin states.

\section{\label{sec:level1}State-Channel duality applied to Standard Model scattering amplitudes}

In quantum information theory, the initial and final states of a qudit are connected through operations called quantum channels. The question we address in this section is whether we can find quantum channels connecting the density matrices describing the initial and final states of particle physics processes. Our proposal is that the Choi-Jamiolkowski isomorphism, also known as state-channel duality, allows us to give an affirmative answer and to apply systematically many more quantum information procedures to particle scattering (cf. Sections V-VII below).\\

However, before we proceed to show how it is applied to particle physics, it is worth noticing that the expression “Choi-Jamiolkowski isomorphism” is often indistinctly used to refer to three distinct isomorphisms: between a general map and a general operator; between a positive map and a positive operator; and finally between a completely positive map and a positive operator\cite{PhysRevA.87.022310}. The latter is the most used and explored isomorphism, and establishes that any channel, by which we mean one or more quantum operations (i.e.  $\mathcal{B}(\mathcal{H})$ is the bounded operator $\mathcal{B}$ on Hilbert space $\mathcal{H}$) corresponding to a linear, completely positive, trace preserving map $\mathcal{E}: \mathcal{B}(\mathcal{H}) \rightarrow \mathcal{B}\left(\mathcal{H}^{\prime}\right)$, connecting an input state space $\mathcal{H}$ of dimension $d$ and an output state space $\mathcal{H}^{\prime}$ of dimension $d^{\prime}$ corresponds to a bipartite state of the tensor product of the two. 

To show this, we follow the exposition in ref.\cite{ekert_2022}. First, we pick $d^{2}$-many basis matrices $|i\rangle\langle j|$ of $\mathcal{B}(\mathcal{H})$, where $i, j \in\{1,2 \ldots, d\}$ and use them to characterise the density operator $\hat{\rho}$ on which the channel $\mathcal{E}$ acts:

\begin{equation}
\mathcal{E}(\hat{\rho})=\mathcal{E}\left(\sum_{i j} \hat{\rho}_{i j}|i\rangle\langle j|\right)=\sum_{i} \hat{\rho}_{i j} \mathcal{E}(|i\rangle\langle j|).
\end{equation}

\noindent We then tabulate all the possible $\left(d^{\prime} \times d^{\prime}\right)$ matrices $\mathcal{E}(|i\rangle\langle j|)$ in $\mathcal{H}^{\prime}$ by constructing the $\left(d d^{\prime} \times d d^{\prime}\right)$ block matrix $\mathcal{H} \otimes \mathcal{H}^{\prime}$, which is called the Choi matrix $\tilde{\mathcal{E}}$:

\begin{equation}
\widetilde{\mathcal{E}}=\frac{1}{d}\left[\begin{array}{cccc}
\mathcal{E}(|0\rangle\langle 0|) & \mathcal{E}(|0\rangle\langle 1|) & \mathcal{E}(|0\rangle\langle 2|) & \cdots \\
\mathcal{E}(|1\rangle\langle 0|) & \mathcal{E}(|1\rangle\langle 1|) & \mathcal{E}(|1\rangle\langle 0|) & \cdots \\
\mathcal{E}(|2\rangle\langle 0|) & \mathcal{E}(|2\rangle\langle 1|) & \mathcal{E}(|2\rangle\langle 2|) & \cdots \\
\vdots & \vdots & \vdots & \ddots
\end{array}\right]
\end{equation}

\noindent The above matrix $\tilde{\mathcal{E}}$ is just an equivalent representation of the quantum channel $\mathcal{E}: \mathcal{B}(\mathcal{H}) \rightarrow \mathcal{B}\left(\mathcal{H}^{\prime}\right)$. \\

In quantum computing, state channel duality is mainly used to assess whether a given channel is physically realisable by considering whether the system’s Choi matrix is a density matrix (i.e. positive: $\mathcal{E}(\rho) \geqslant 0$ whenever $\rho \geqslant 0$ and trace preserving: $\operatorname{tr} \mathcal{E}(\rho)=\operatorname{tr} \rho$ for all $\rho$). It is worth noting, however, that because the operator-sum (Kraus) representation of a given channel is not unique\cite{ekert_2022}, different channels can lead to the same density matrix. Hence, the state-channel duality allows one to ascertain whether a given channel can be physically implemented, but it does not imply that this will be the only possible implementation of a given density matrix.

Our suggestion is to apply this to high energy scattering processes by making the reverse use of the duality: because through quantum state tomography\cite{ashbypickering2022quantum, Barr_2022} it is possible to reconstruct the density matrix of a given process, which will also automatically fulfil the formal characteristics required for it to correspond to a physical process, such as positivity\cite{Martens_2017} ($\rho \geqslant 0$), we conclude that it is in principle possible to construct a quantum channel connecting the initial and final state density matrices. \\

\section{\label{sec:level2}An example model: $\mathrm{e}^{+} \mathrm{e}^{-} \rightarrow t \bar{t}$  electroweak process with polarised beams}

In this section we present the process of electron-positron annihilation producing a top quark and its anti-quark with polarised beams as an example model to illustrate how state-channel duality may be applied to simulate helicity states density matrices. 

This process is particularly interesting because the leptonic decays of the $t$ and $\bar{t}$ (anti-) quarks effectively induce a measurement of spin along the axis of the emitted lepton\cite{Fabbrichesi_2021}, thus allowing access to the spin states of the $t \bar{t}$ system and making the quantum-computing simulated density matrix amenable to experimental verification. Hence, it allows us to test the assumption that colliders can be identified as privileged observatories for quantum information processing (cf. Section I, point (i)). Other Standard Model processes for which spin tomography is possible include the decay of $W$ bosons (e.g. produced from Higgs boson decays) to leptons\cite{Barr_2022}.

\begin{figure}[htp]
    \centering
    \includegraphics[width=9cm]{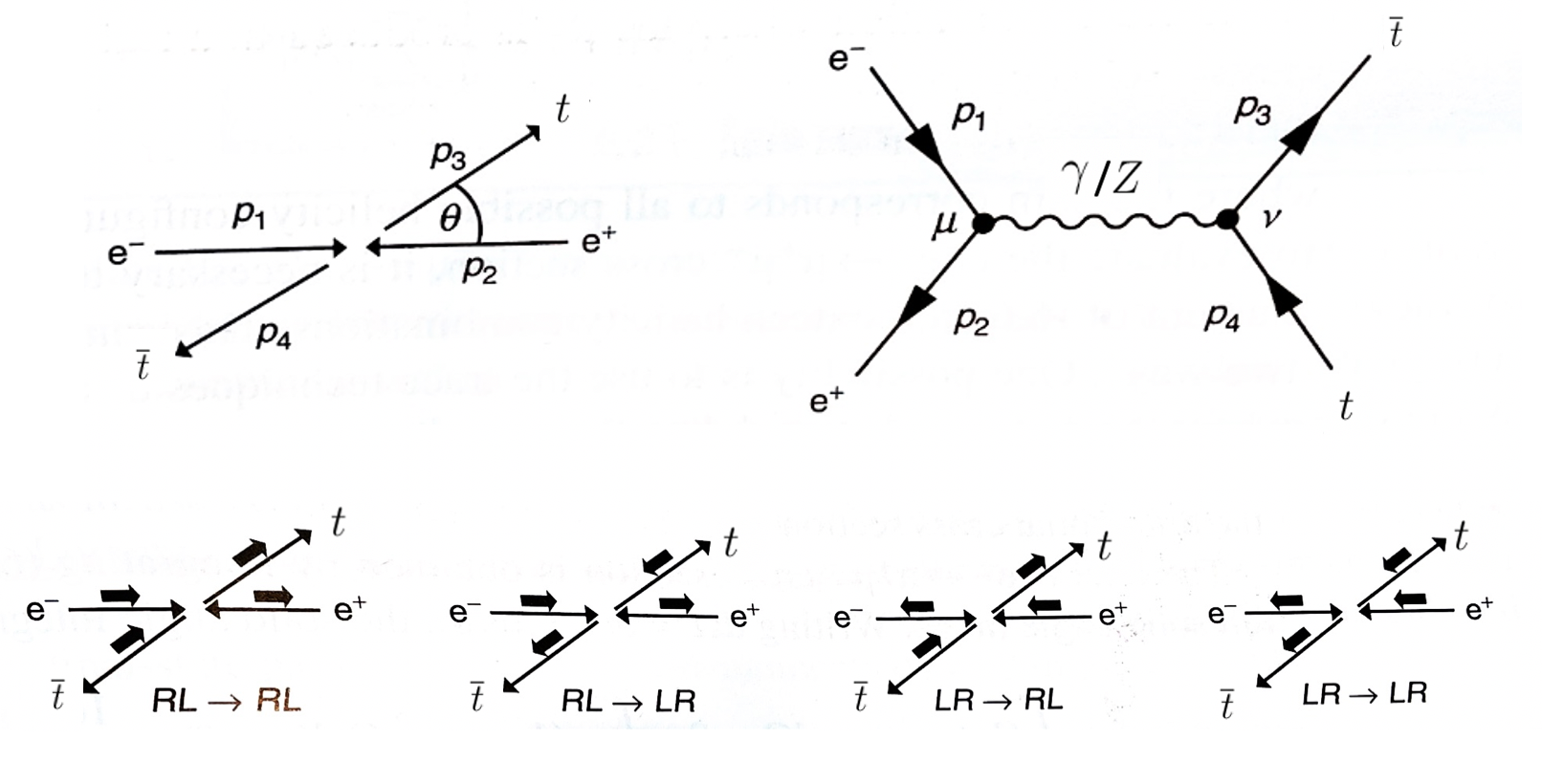}
    \caption{The $\mathrm{e}^{+} \mathrm{e}^{-} \rightarrow t \bar{t}$ electroweak process viewed in the centre-of-mass frame (top-left), corresponding lowest-order Feynman diagram (top-right), and four of the $16$ possible helicity combinations (adapted from ref.\cite{thomson_2021})}
    \label{}
\end{figure}

\subsection{Notes on polarisation choices}

Our example process $\mathrm{e}^{+} \mathrm{e}^{-} \rightarrow t \bar{t}$ contains two spin-$\frac{1}{2}$ particles in the incoming state, and two spin-$\frac{1}{2}$ particles in the outgoing state. Thus both the initial and final state are naturally represented by two-qubit systems, and the Hilbert state for the initial state (and similarly that for the final state) is therefore of dimension $2^2=4$. The density operators on each of those Hilbert spaces are $4\times4$ matrices, and the full Choi matrix, which maps the initial to the final state (eq. 6), is of dimension $16 \times 16$.  

To more simply illustrate the principles we consider a reduced subset of the Choi matrix elements relating particular polarisation choices of the initial and final states. We restrict our attention to those initial- and final-state density matrices that can be represented by polarisation (Bloch) vectors that point along the momentum direction of each particle, which is equivalent to saying that the states that are diagonal in the helicity basis. Since the density matrices are diagonal in the helicity basis the Choi matrix elements representing them are also diagonal. 

This restriction should not be interpreted as meaning that the other Choi elements, those that are off-diagonal in the helicity basis, are necessarily inaccessible. For the final state, techniques have been developed \cite{Afik_2021, ashbypickering2022quantum} that can reconstruct the full bipartite density matrix for a pair of top quarks, so all possible densities in the final-state Hilbert space can be accessed and measured experimentally. For the initial state, technologies exist (see e.g. ref.\cite{Sokolov:1963zn,sobloher2012polarisation, blondel2019polarization}) that allow us to polarise high-energy electron beams in both the longitudinal direction (i.e. with the spin polarisation vector (anti-)parallel to the beam) and in the transverse direction, thus each beam having an arbitrary Bloch vector. The beams are not entangled, so the density matrices of the bipartite initial state which are experimentally accessible are those that can be represented by the product state

\begin{equation}
\rho=\rho_A \otimes \rho_B
\end{equation}

\noindent for some $\rho_A$ and $\rho_B$ representing the density matrices of the two beams respectively.

In what follows, we restrict our attention to only those Choi matrix elements that are diagonal in the helicity basis in order to more compactly illustrate the technique. For compactness, we report these values in a $4 \times 4$ matrix, with the initial state polarisations represented by rows and the final-state polarisations represented by columns.

\subsection{Example model illustration}

Following the traditional quantum field theoretical approach (cf. ref.\cite{thomson_2021} for full steps), the process cross-section $\sigma$ at leading order in the electroweak coupling constants is proportional to

\begin{equation}
\sigma \propto|\mathcal{M}|^2=\left|\mathcal{M}_\gamma+\mathcal{M}_Z\right|^2
\end{equation}

\noindent where $\mathcal{M}_\gamma$ and $\mathcal{M}_Z$ are the density matrices related to the $Z$ boson and photon, $\gamma$, exchange Feynman diagrams (Fig. 1). Since our points are sufficiently illustrated by considering separately the cases of the photon and $Z$ boson propagators, we neglect in what follows the interference term $2\left|\mathcal{M}_\gamma \mathcal{M}_Z{ }^*\right|$. Adding that term is a straight-forward modification, but it adds to the complexity of the expression and is not required for our illustrative purposes.

Let us first consider the individually electromagnetic process associated with $\mathcal{M}_\gamma$, which is given by

\DIFaddend \begin{equation}
\DIFdelbegin 
\DIFdelend \DIFaddbegin \begin{aligned}
\mathcal{M}_\gamma &=-\frac{e^2}{q^2} g_{\mu v}\left[\bar{v}\left(p_2\right) \gamma^\mu u\left(p_1\right)\right]\left[\bar{u}\left(p_3\right) \gamma^\nu v\left(p_4\right)\right] \\
&=-\frac{e^2}{s} j_e \cdot j_t,
\end{aligned}\DIFaddend 
\end{equation}

\noindent where $\bar{v}\left(p_2\right)$, $u\left(p_1\right)$, $\bar{u}\left(p_3\right)$, $v\left(p_4\right)$ are the helicity spinors associated with the initial-state positron, initial-state electron, final-state top quark and final-state antitop quark respectively, $\gamma^\mu$ are the Dirac–Pauli representation of the $\gamma$-matrices, $e^2/4\pi$ is the electromagnetic fine-structure constant, $g_{\mu v}$ is the Minkowski metric, $q$ is the four-momentum of the photon propagator, $j_e$ and $j_t$ are the four-vector electron-postron and top-antitop currents respectively, and the centre-of-mass energy $\sqrt{s}=2 E_{\text {beam }}$ is equal to twice the beam energy.

Because we do not wish to average over spin degrees of freedom, we need to obtain and multiply together the electron-positron and top-antitop currents for every one of the possible 16 helicity combinations (e.g. $t_{\uparrow} \bar{t}_{\downarrow}=j_{t, R L}=\left( j_{t}^{0}, j_{t}^{1}, j_{t}^{2}, j_{t}^{3}\right)$) without using spin trace techniques. We thus construct the matrix 

\begin{equation}
\left|\mathcal{M}_\gamma\right|^2 \propto\left[\begin{array}{cccc}
\left|M_{R L \rightarrow R L}\right|^2 & \left|M_{R L \rightarrow R R}\right|^2 & \left|M_{R L \rightarrow L L}\right|^2 & \left|M_{R L \rightarrow L R}\right|^2 \\
\left|M_{R R \rightarrow R L}\right|^2 & \left|M_{R R \rightarrow R R}\right|^2 & \left|M_{R R \rightarrow L L}\right|^2 & \left|M_{R R \rightarrow L R}\right|^2 \\
\left|M_{L L \rightarrow R L}\right|^2 & \left|M_{L L \rightarrow R R}\right|^2 & \left|M_{L L \rightarrow L L}\right|^2 & \left|M_{L L \rightarrow L R}\right|^2 \\
\left|M_{L R \rightarrow R L}\right|^2 & \left|M_{L R \rightarrow R R}\right|^2 & \left|M_{L R \rightarrow L L}\right|^2 & \left|M_{L R \rightarrow L R}\right|^2
\end{array}\right]\DIFaddbegin \DIFadd{.
}\DIFaddend \end{equation}

\noindent{Disregarding any pre-factors, we obtain the following matrix}

\begin{equation}
\left|\mathcal{M}_\gamma\right|^2 \propto\left[\begin{array}{cccc}
(1+\cos \theta)^2 & 0 & 0 & (1-\cos \theta)^2 \\
(\sin \theta)^2 & 0 & 0 & (\sin \theta)^2 \\
(\sin \theta)^2 & 0 & 0 & (\sin \theta)^2 \\
(1-\cos \theta)^2 & 0 & 0 & (1+\cos \theta)^2
\end{array}\right]\DIFaddbegin \DIFadd{.
}\DIFaddend \end{equation}

On the other hand, in the limit where chiral and helicity states are essentially the same (ultra-relativistic),  $\mathcal{M}_Z$ can be directly derived in a similar manner to (7) above, but introducing the couplings between the $Z$ boson and the fermions involved, and noticing that some helicity combinations give $zero$ matrix elements. One finds

\begin{equation}
\left|\mathcal{M}_Z\right|^2 \propto\left[\begin{array}{cccc}
\left(c_R^e c_R^t\right)^2(1+\cos \theta)^2 & 0 & 0 & \left(c_R^e c_L^t\right)^2(1-\cos \theta)^2 \\
0 & 0 & 0 & 0 \\
0 & 0 & 0 & 0 \\
\left(c_L^e c_R^t\right)^2(1-\cos \theta)^2 & 0 & 0 & \left(c_L^e c_L^t\right)^2(1+\cos \theta)^2
\end{array}\right]
\end{equation}

\noindent where $c_L$ and $c_R$ are the coupling constants of the left-chiral and right-chiral fermions respectively. These calculations already remind one of the construction of a Choi matrix (eq. 2), and we can easily check that the obtained matrices satisfy the conditions to be implemented through quantum channels (cf. Section III). In fact, using the IBM Quantum Lab (https://quantum-computing.ibm.com/), we constructed quantum channels that were able to reproduce them, which is the subject of the next section.

\subsection{\label{sec:level2} CONSTRUCTING THE QUANTUM CHANNELS}

When constructing a quantum channel (Fig. 2), we translate observable values R, L to the Boolean computational basis given by \{0,1\}. We can characterise each possible initial-final helicity state by a string of four digits, the first two corresponding to the helicity of the initial-state electron and positron, and the last two corresponding to the helicity of the final-state quarks. The first application of the Hadamard gate (H) creates an equal superposition of all the 16 possible combinations of initial-final helicity states. Then through the quantum interference introduced by the action of phase and C-NOT gates we enhance the probability amplitudes of selected states. The final Hadamard gates `close' the superposition and precede the measurement gates. It is worth noting that quantum channels can approximate a result with as much precision as needed, at the cost of increasing the number of computational steps (cf. Solovay-Kitaev theorem \cite{nielsen2010quantum}). \\

\begin{figure}[htp]
    \centering
    \includegraphics[width=8cm]{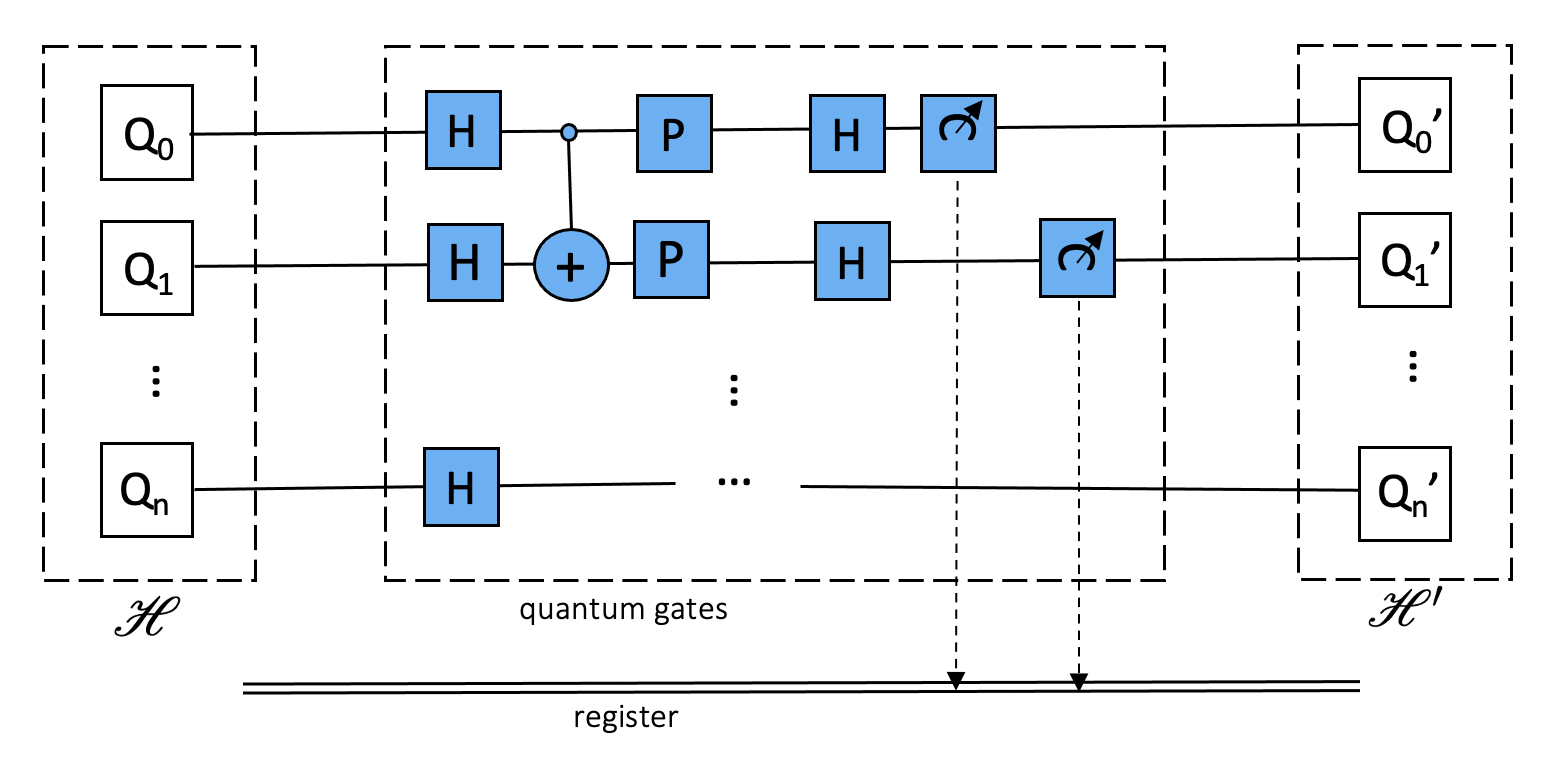}
    \caption{Schematic representation of a quantum channel comprising $n$ qubits, Hadamard (H), C-NOT ($+$), Phase (P), and measurement gates.}
    \label{}
\end{figure}

\begin{figure}[htp]
    \centering
    \includegraphics[width=9cm]{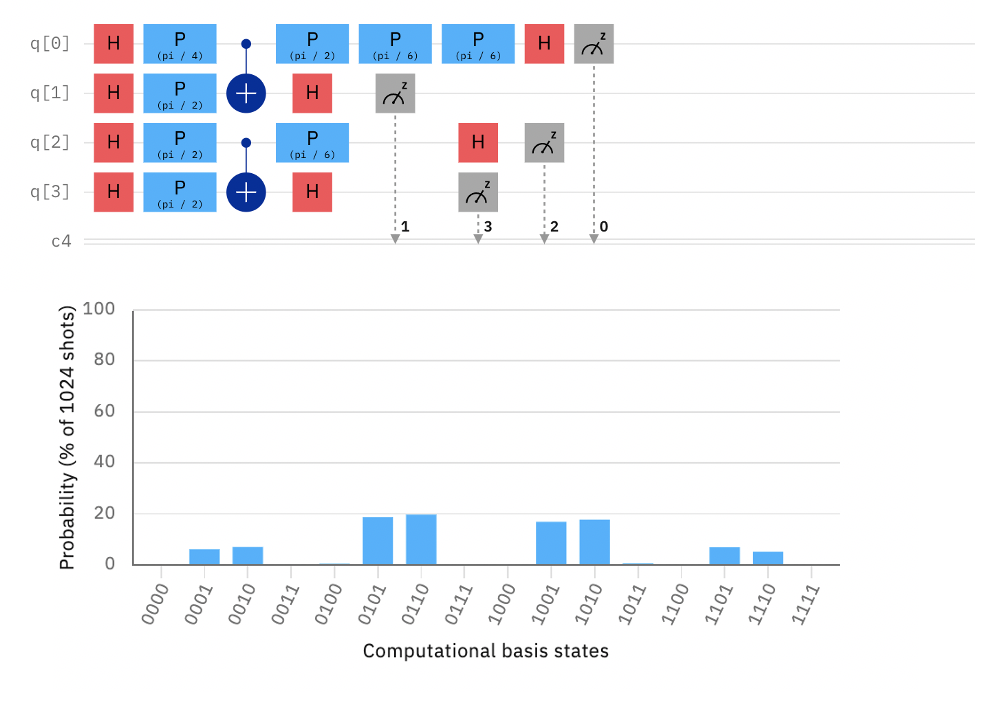}
    \caption{Quantum channel simulating $|\mathcal{M}_\gamma|^2$ (eq. 7) and related histograms showing a parity preserving process (simulated using the IMB Quantum Lab (https://quantum-computing.ibm.com). }
    \label{}
\end{figure}

\begin{figure}[htp]
    \centering
    \includegraphics[width=9cm]{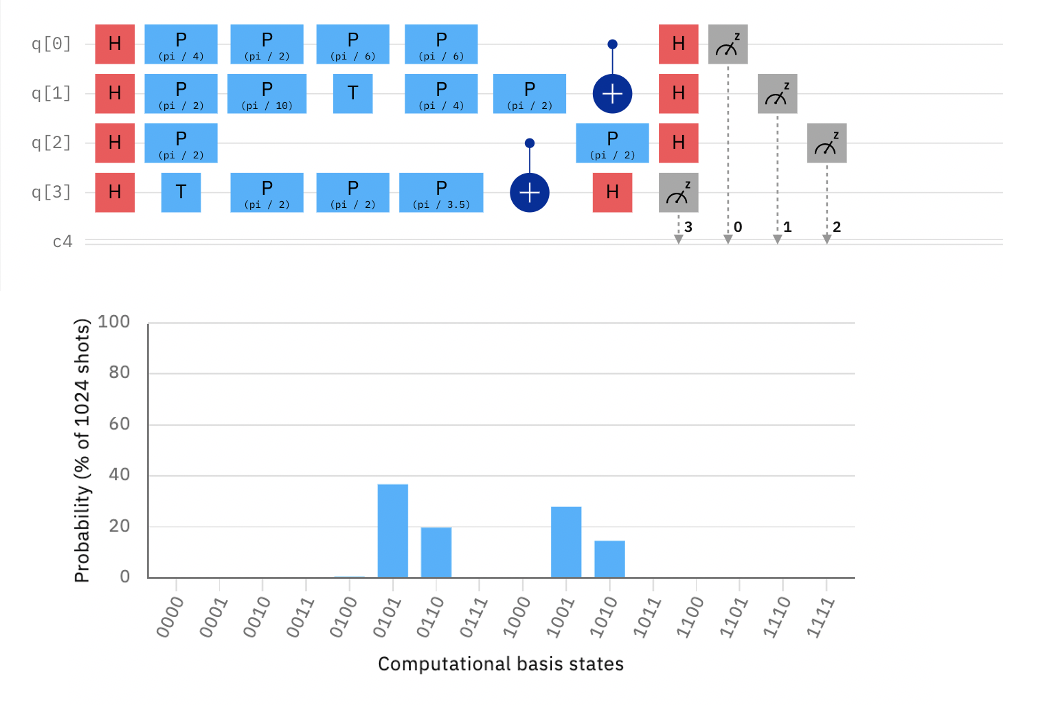}
    \caption{Quantum channel simulating $|\mathcal{M}_Z|^2$ (eq. 8) and related histograms showing a parity violating process (simulated using the IMB Quantum Lab (https://quantum-computing.ibm.com).}
    \label{}
\end{figure}

Considering the experimental measurements of the $\mathrm{e}^{+} \mathrm{e}^{-} \rightarrow q \bar{q}$ (where $q$ and $\bar{q}$ are a quark and its anti-quark respectively) cross section from LEP close to and above $Z$ resonance (from LEP and SLD Collaborations
\cite{Physics_Reports_2006}, we see that the contribution to the cross section from the QED process alone (eq. 7) presents an evident backward-forward symmetry in the angular distribution that is lost above the $Z$ resonance as a consequence of the parity violation which characterises the weak interaction (eq. 8), that is manifested via different couplings for the left- and right-chiral fermions. Figures 3-4 illustrate a simple example of how quantum circuits can be used to compute ansatz leading-order matrix elements for $|\mathcal{M}_\gamma|^2$ and $|\mathcal{M}_Z|^2$, which respect the observed angular distributions (a)symmetries. 

Consider, for example, how the histogram in Fig. 3 is associated with the elements in matrix (7). A given combination of 0's and 1's determines a unique matrix element according to matrix (6) (as an example, the top-left matrix element $\left|M_{R L \rightarrow R L}\right|^2$ is associated with the combination "0101"), while the associated probability amplitude gives the corresponding value according to matrix (7). The parity preserving nature of the electric interaction (eq. 7) can be seen in Fig. 3 by the fact that, by substituting each 0 with a 1 and vice-versa in each string, the same graph would be obtained.
On the other hand, in the weak interaction (eq. 8) parity violation manifests itself via the unequal values of the chiral couplings $c_L$ and $c_R$. It follows that, by comparing matrix (8) with matrix (7), we expect that matrix elements that had equal values in the former, will be unequal in the latter. This is what we observe in Fig. 4, where we also notice that substituting each 0 with a 1 and vice-versa in each string would not lead to the same graph. Adding the interference term $2\left|\mathcal{M}_\gamma \mathcal{M}_Z{ }^*\right|$ would similarly lead to parity violation at and above the $Z^0$ boson resonance. \\

In both cases the obtained probabilities are in agreement with the theoretical predictions of eqs. (4-8). It is worth noting that, while the data we used to construct the channels in the presented example model are not experimental, there are no barriers to using experimental data in this context. Given the quantity and high precision of data produced at the LHC and other colliders, this provides an interesting research direction. The quantum circuits presented could in fact be optimised by employing, for example, a hybrid quantum-classical variational approach, where the effect of circuit gates is modified to match a specific experimental distribution by tunable parameters optimised by classical computing techniques (cf. Section VII for possible interpretations of a mismatch between experimental and quantum channel simulated distributions). Such techniques have been successfully applied to data analysis at colliders: two recent proposals involve using unsupervised training of generative models to generate synthetic data of high-energy physics processes
\cite{https://doi.org/10.48550/arxiv.2203.03578}, and the quantum variational classifier method to recognise signatures from experimental data from LHC\cite{Wu_2021}. These methods share a common structure, as they consist in repeatedly producing synthetic ansatz density matrices that are then compared with experimental data and updated.

 In this Section we have illustrated how state-channel duality can be applied in the context of a simple $2\rightarrow2$ particle process when considered at the leading order in perturbation theory. In the following Sections we describe how this could be generalised to additional processes and higher orders in perturbation theory.

\section{\label{sec:level1} CONSTRUCTING A DICTIONARY}

We suggest that some useful quantum channel operations, such as channel concatenation and factorisation, which follow as corollaries of state-channel duality\cite{https://doi.org/10.48550/arxiv.1509.08339}, can be associated with a physical meaning in the field of particle physics. For the quantum information nomenclature in this section, we follow ref.\cite{wilde_2017}. \\
\\
\textit{a. Serial concatenation} submits the quantum state to subsequent quantum channels. Let $\mathcal{N}: \mathcal{B}(\mathcal{H}_A) \rightarrow \mathcal{B}\left(\mathcal{H}_B\right)$ and $\mathcal{M}: \mathcal{B}(\mathcal{H}_B) \rightarrow \mathcal{B}\left(\mathcal{H}_C\right)$ correspond to the first and second quantum channels respectively, with the respective Kraus operators being $\left\{\mathrm{N}_{\mathrm{k}}\right\}$ and $\left\{\mathrm{M}_{\mathrm{k}}\right\}$. The output of the first channel $\mathcal{N}_{A \rightarrow B}\left(\rho_A\right) \equiv \sum_k N_k \rho_A N_k^{\dagger}$ is submitted to the second thus giving:

\begin{equation}
\left(\mathcal{M}_{B \rightarrow C} \circ \mathcal{N}_{A \rightarrow B}\right)\left(\rho_A\right)=\sum_k M_k \mathcal{N}_{A \rightarrow B}(\rho) M_k^{\dagger}=\sum_{k, k^{\prime}} M_k N_{k^{\prime}} \rho_A N_{k^{\prime}}^{\dagger} M_k^{\dagger}
\end{equation}
In the context of particle physics, this is applicable to modelling amplitudes for sequential processes such as cascade decays and jets (cf. eq. 1 in ref.\cite{knowles_1990}, and Section VI below) (Fig. 5-a). Notice that the dimensions of the final states Hilbert space need not match that of the initial state: an appending channel can be used to encode the additional degrees of freedom of decay products (cf. point d. below). \\

\textit{b. Parallel concatenation} sends a system $A$ through a channel $\mathcal{N}: \mathcal{B}(\mathcal{H}_A) \rightarrow \mathcal{B}\left(\mathcal{H}_C\right)$ and a system $B$ through a channel $\mathcal{M}: \mathcal{B}(\mathcal{H}_B) \rightarrow \mathcal{B}\left(\mathcal{H}_D\right)$.
  Thus, for an input density operator $\rho_{A B} \in \mathcal{D}\left(\mathcal{H}_A \otimes \mathcal{H}_B\right)$, we have: 

\begin{equation}
\left(\mathcal{N}_{A \rightarrow C} \otimes \mathcal{M}_{B \rightarrow D}\right)\left(\rho_{A B}\right)=\sum_{k, k^{\prime}}\left(N_k \otimes M_{k^{\prime}}\right)\left(\rho_{A B}\right)\left(N_k \otimes M_{k^{\prime}}\right)^{\dagger}
\end{equation}

\noindent In the context of particle physics this is applicable where subsystems are present and each subsystem undergoes such a process that can be regarded as independent of the state of the other subsystem, i.e. it does not introduce entanglement. As an example, consider parallel decays (Fig. 5-b). \\
\\
\textit{c. A preparation channel} prepares a quantum system $A$ in a given state $\rho_{A} \in \mathcal{D}(\mathcal{H}_A)$, and can correspond, for example, to the state of one of the two incoming beams in Section IV. In Fig. 3-4 above, the $q[0]$ and $q[1]$ registers, encoding the helicity states of the incoming electron and positron respectively, are prepared in an equal superposition of all initial helicity states by applying an Hadamard gate ($H$) on each register (Fig. 5-c). \\
\\
\textit{d. An appending channel} is the parallel concatenation of the identity channel and a preparation channel, and can correspond, for example, to the state of second of the two incoming beams in Section IV (Fig. 5-d), as it can be noticed by the fact that in the simulations leading to  Figs. 3 \& 4 above each register was individually prepared and appended to each other.\\

\begin{figure}[htp]
    \centering
    \includegraphics[width=9cm]{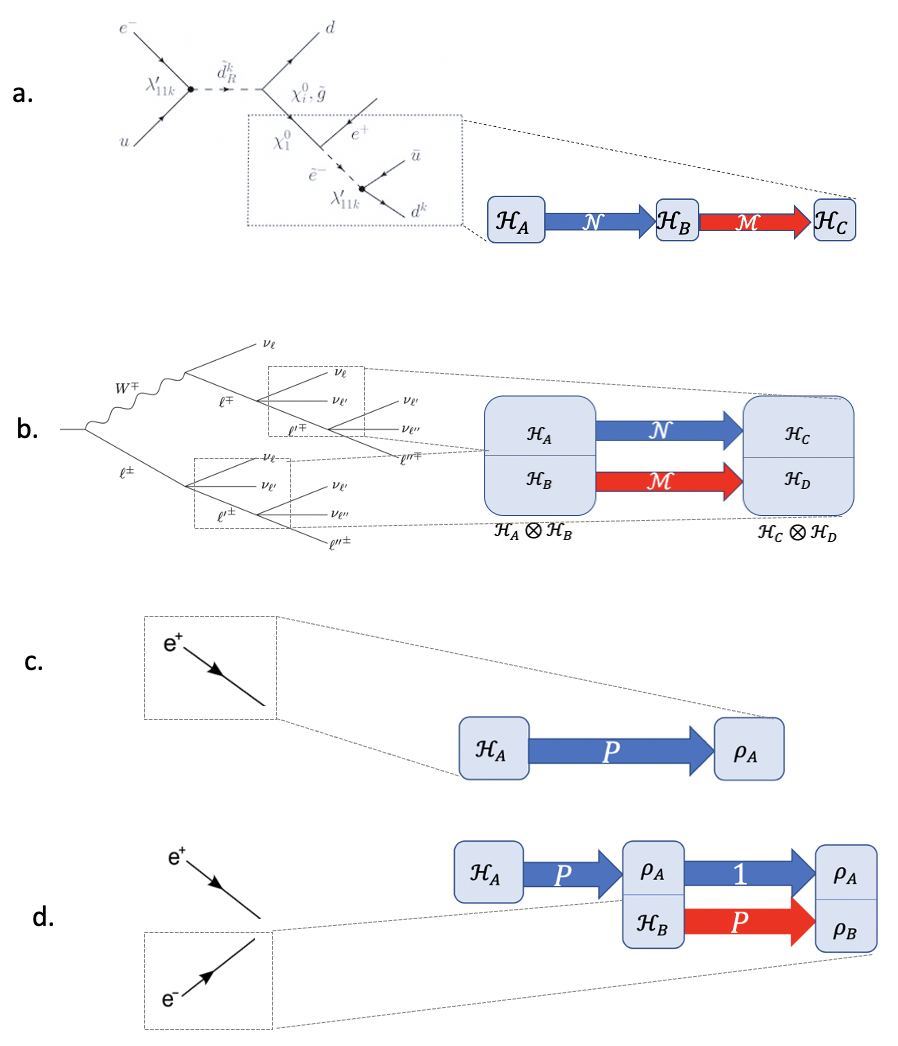}
    \caption{Examples from the text of:  (a) serial concatenation of channels, (b) parallel concatenation of channels, (c) a preparation channel, (d) an appending channels and their suggested interpretation in terms of particle physics processes.}
    \label{}
\end{figure}

\section{EXTENSIONS OF THE PRESENT WORK: higher-order corrections}

While in the previous sections we applied state-channel duality to amplitudes corresponding to tree-level Feynman diagrams, we expect that a mapping is also possible at higher orders in perturbation theory, where quantum corrections are included. 

Consider the Feynman path integral formulation
\cite{RevModPhys.20.367}, which relates the quantum field theoretic picture (RHS) to a quantum mechanical amplitude (LHS)

\begin{equation}
\left\langle\Phi\left(t_f\right) \mid \Phi\left(t_i\right)\right\rangle=\int \mathcal{D}[\Phi] \mathrm{e}^{i \mathcal{S}[\Phi]}
\end{equation}

\noindent where $\mathcal{S}$[$\Phi$] is the action, functionally dependent on field configurations $\Phi$, and $\mathcal{D}[\Phi]$ is the measure of integration. Notice that the LHS of eq. 11 above is a quantum mechanical amplitude, which makes it suitable for the application of state-channel duality. We hence expect, within the limits of applicability of eq. 11, to be able to find an equivalent unitary operator (or collection thereof) corresponding to a quantum channel.\\

We can identify such operator(s) by considering the equivalent S-matrix formulation. Following the exposition in the classical ref.\cite{peskin_schroeder_2019}, according to this formulation the transition amplitude between input and output states of definite momentum is given by 

\begin{equation}
_{\text {out }}\left\langle\mathbf{p}_1 \mathbf{p}_2 \cdots \mid \mathbf{k}_{\mathcal{A}} \mathbf{k}_{\mathcal{B}}\right\rangle_{\text {in }} \equiv\left\langle\mathbf{p}_1 \mathbf{p}_2 \cdots|S| \mathbf{k}_{\mathcal{A}} \mathbf{k}_{\mathcal{B}}\right\rangle
\end{equation}

\noindent where $S$ is a unitary operator, i.e. $S^{\dagger} S=1$. It is worth noting that in quantum field theory, $S$ is also \textit{limiting unitary}, which means that it can be decomposed in such a way that \textit{`in and out states are related by the limit of a sequence of unitary operators'}\cite{peskin_schroeder_2019}. 

Following a standard procedure (cf. ref.\cite{peskin_schroeder_2019}, Section 4.5, and specifically eq. 4.73 therein), it is possible to see that this condition remains valid at any order in perturbation theory. $S$ can in fact be decomposed into

\begin{equation}
S = \mathbf{1} + iT
\end{equation}

\noindent where the $T$ matrix encodes the interactions and is proportional to the unitary operator 

\begin{equation}
U\left(t, t_0\right)=
T\left\{\exp \left[-i \int_{t_0}^t d t^{\prime} H_I\left(t^{\prime}\right)\right]\right\}
\end{equation}

\noindent where $H_I$ is the interacting Hamiltonian proportional to a small coupling constant. Higher-order corrections in perturbation theory are calculated through the power series expansion of the $T$ operator, and amplitudes at each order tolerate an error that is proportional to the related power in the coupling constant. \\

Because of the unitarity of the $S$ matrix, the sum of all the relevant contributions associated with it (eq. 12 corresponds to the sum over all connected, amputated Feynman diagrams with the specified number of fields and vertices) up to a given order in perturbation theory has the form of a quantum mechanical amplitude, and can thus be interpreted as a quantum channel. Also notice that because it includes all the possible processes, this is the relevant amplitude when performing unitarity tests (cf. section VII below for discussion).\\

Let us illustrate the above points by considering the radiative corrections in Fig. 6. 

\begin{figure}[htp]
    \centering
    \includegraphics[width=9cm]{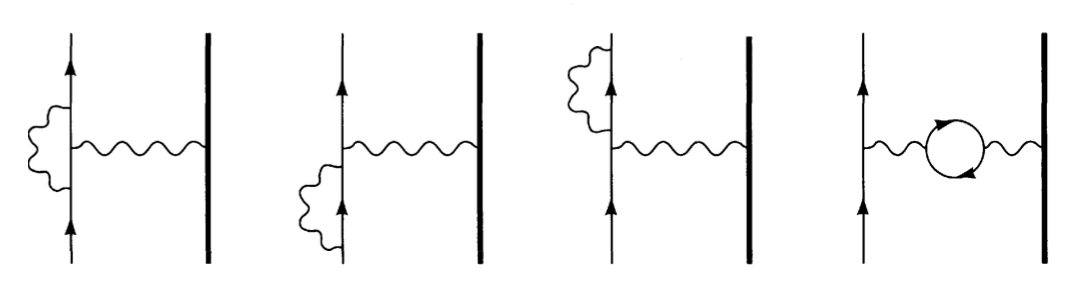}
    \caption{Feynman diagrams displaying vertex, external leg and vacuum polarisation corrections respectively (adapted from ref.\cite{peskin_schroeder_2019}).}
    \label{}
\end{figure}

\begin{figure}[htp]
    \centering
    \includegraphics[width=6cm]{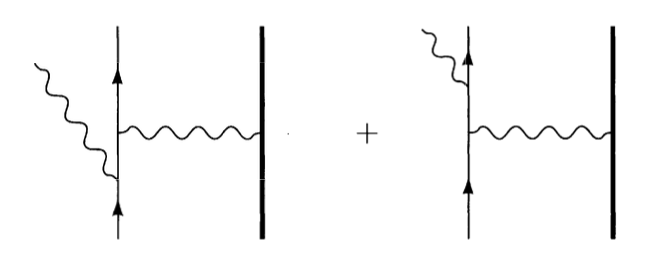}
    \caption{Bremsstrahlung diagrams (adapted from ref.\cite{peskin_schroeder_2019})}
    \label{}
\end{figure}

\noindent While we refer to the classical ref
\cite{peskin_schroeder_2019} (Sections 7.2-3)
for a detailed calculation of each contribution, for the purpose of the present argument it suffices to notice that the inclusion of all the relevant corrections (fig. 6-7) preserves the quantum-mechanical amplitude nature of the S-matrix (cf. eq. 12), since ultraviolet divergences cancel out of expressions for observable quantities such as cross sections, and the infrared divergences associated to the diagrams in Fig. 6 are cancelled out by adding the contributions due to the soft Bremsstrahlung (fig. 7), thus allowing for the application of state-channel duality to the amplitude corresponding to the sum of all the relevant corrections.\\

Additionally, there exist cases when the application of state-channel duality to specific corrections is also possible. As an example, in cases where the correction corresponds to the inclusion of an additional final-state particle, and calculations allow for the factorisation of the tree-level result, the channel composition rules from Section V can applied. As a concrete example, consider the amplitude associated to the Bremsstrahlung process (fig. 7):
 
\begin{equation}
i M=\bar{u}\left(p^{\prime}\right)\left[M_0\left(p^{\prime}, p\right)\right] u(p) \cdot\left[e\left(\frac{p^{\prime} \cdot \varepsilon^*}{p^{\prime} \cdot k}-\frac{p \cdot \varepsilon^*}{p \cdot k}\right)\right],
\end{equation}

\noindent which is the amplitude for tree-level elastic scattering multiplied by an extra factor accounting for the emission of the photon (where $k$ and $\varepsilon^*$ are the photon momentum and polarisation vector respectively) (cf. Section 6.1 of ref.\cite{peskin_schroeder_2019}, and specifically eq. 6.22 therein). On the quantum computing side, this could be modelled as a quantum channel corresponding to the amplitude for tree-level elastic scattering, followed by the application, through serial concatenation, of an `amplitude damping' channel, that accounts for the extra factor. This operation effectively embeds the density matrix associated to the lower-order Feynman diagram into that of the higher-order one by rescaling the relevant degrees of freedom (thus `damping it') to account for the additional final-state degrees of freedom. Notice this preserves the unitarity of the overall process once the degrees of freedom of the additional final-state photon are also taken into account, and has elsewhere\cite{preskill_2018} been used to model spontaneous emission processes.\\

 \section{FINAL REMARKS}

State-channel duality may prove to be a useful instrument for foundational tests of quantum mechanics at colliders as it allows us to tackle rarely explored features of high energy physics systems by analysing their corresponding quantum channels. By directing our attention not only to a given density matrix, but also to the corresponding quantum channels, we may be able to systematically search for properties such as entanglement, rotation invariant quantities and symmetries. As an example, separability considerations related to state-channel duality\cite{antipin2020channelstate} could be applied to probe the quantum mechanics features in particle physics problems, even at leading order (cf., e.g., ref.\cite{eckstein_horodecki_2022}). In fact, since \emph{"bipartite states are well classified according to correlations therein \cite{PhysRevA.40.4277,Horodecki_2009,PhysRevA.77.022301}, we may classify channels by translating the classifications in bipartite states with the mediation of channel-state duality and investigate the conditions for channels in order to generate corresponding states with certain correlation structures."}\cite{PhysRevA.87.022310}.\\

In addition, it is worth noting that effective field theories that respect the limiting unitarity condition outlined in Section VI can be modelled as quantum channels though `oracle operators'\cite{nielsen2010quantum} connecting input to output states. This may have an application, where physics beyond the Standard Model is expected to be observed, as experimentally probing all matrix elements through polarisation choices (cf. Section IV-A) would allow one to construct a theory-independent quantum channel, from which it could nonetheless be possible to extract information on the underlying theory such as chirality structure and various forms of symmetry violation. In fact, since state-channel duality establishes a mapping between the quantum channel and the S-matrix, and not directly the Hamiltonian, the quantum channel formulation can be agnostic to the underlying full particle physics Hamiltonian while outputting a result that is nonetheless compatible with the underlying dynamics. 

Finally, notice that when an experimentally measured density matrix cannot be turned into a quantum channel, this might be taken to suggest that it cannot be produced by unitary dynamics alone and, if the traditional quantum mechanics picture is complete and the perturbative approach applicable at given energy level, then additional interactions (e.g. with a subsystem) may be present.\\

For these reasons, we are hopeful the presented framework will prove to be a useful tool for the particle physics community, and provide an interesting addition to the quantum computing methods applied to particle physics problems
\cite{https://doi.org/10.48550/arxiv.2203.03578,Wu_2021,PhysRevD.106.056002}.

\section{\label{sec:level1} ACKNOWLEDGEMENTS}

CA is particularly grateful to Malcolm Fairbairn for helpful comments and discussion. We would like to thank the members of the Oxford ATLAS ‘SM \& beyond’ group, the participants to the conferences ‘Foundational Tests of Quantum Mechanics at the LHC’ (Merton College, Oxford, 20-22 March 2023) and ‘Quantum Computing for High-Energy Physics’ (Durham, 19-20 September 2023) and particularly Michał Eckstein, for enlightening discussion. We would also like to thank two anonymous referees for valuable comments and suggestions that helped us improve the present work. We acknowledge the use of IBM Quantum services for this work. The views expressed are those of the authors and do not reflect the official policy or position of IBM or the IBM Quantum team. AJB is funded through STFC grants ST/R002444/1 and ST/S000933/1, by the University of Oxford and by Merton College, Oxford. CA is funded through a King's College London NMES studentship. For the purpose of Open Access, the authors have applied a CC BY public copyright licence to any Author Accepted Manuscript (AAM) version arising from this submission.

\appendix

\section{Choice of basis example: helicity}

As an example of choice of basis, the particle (eq. 1) and antiparticle (eq. 2) right ($R$) and left ($L$) handed chiral Dirac spinors satisfying

\begin{equation}
\gamma^5 u_R=+u_R, \;\;\;\;\;\; \gamma^5 u_L=-u_L,
\end{equation}

and

\begin{equation}
\gamma^5 v_R=-v_R,  \;\;\;\;\;\; \gamma^5 v_L=+v_L,
\end{equation}

\noindent being eigenstates of the unitary matrix 

\begin{equation}
\gamma^5 \equiv i \gamma^0 \gamma^1 \gamma^2 \gamma^3=\left(\begin{array}{llll}
0 & 0 & 1 & 0 \\
0 & 0 & 0 & 1 \\
1 & 0 & 0 & 0 \\
0 & 1 & 0 & 0
\end{array}\right)=\left(\begin{array}{ll}
0 & I \\
I & 0
\end{array}\right)
\end{equation}

\noindent provide a good alternative for representing relativistic spin-$\frac{1}{2}$ fermions, as any such state can be prepared on a quantum circuit. We note that generally qubit quantum computers implement unitary operations, e.g. within IBM Quantum lab\cite{IBMQ}, any unitary matrix can be turned into a unitary gate with {\fontfamily{qcr}\selectfont UnitaryGate()} (cf., as a possible protocol, initialising eigenstates of $\gamma^5$, introducing relative phase differences though phase gates corresponding to relative amplitudes, applying a depolarising channel to set all amplitudes to zero, thus obtaining a unitary quantum channel that, when run in reverse, outputs the desired state; cf., for example, ref.\cite{https://doi.org/10.48550/arxiv.1509.08339}).

However, because chirality does not have a straightforward physical interpretation, helicity spinors, which we used in the example model presented in Section IV, 

\begin{equation}
u_{\uparrow}(p)=N\left(\begin{array}{c}
c \\
s e^{i \phi} \\
\frac{\mathrm{p}}{E+m} c \\
\frac{\mathrm{p}}{E+m} s e^{i \phi}
\end{array}\right), u_{\downarrow}(p)=N\left(\begin{array}{c}
-s \\
c e^{i \phi} \\
\frac{\mathrm{p}}{E+m} s \\
-\frac{\mathrm{p}}{E+m} c e^{i \phi}
\end{array}\right),
\end{equation}

\begin{equation}
v_{\uparrow}(p)=N\left(\begin{array}{c}
\frac{\mathrm{p}}{E+m} s \\
-\frac{\mathrm{p}}{E+m} c e^{i \phi} \\
-s \\
c e^{i \phi}
\end{array}\right), v_{\downarrow}(p)=N\left(\begin{array}{c}
\frac{\mathrm{p}}{E+m} c \\
\frac{\mathrm{p}}{E+m} s e^{i \phi} \\
c \\
s e^{i \phi}
\end{array}\right),
\end{equation}

\noindent{(where} $p$ is particle's 3-momentum, $N=\sqrt{E+m}$ 
 is a normalization factor with $E$ and $m$ being particle's energy and mass respectively, $s=\sin \frac{\theta}{2}$, $c=\cos \frac{\theta}{2}$ with azimuthal angle  $\theta$ defined as in Fig.1 and polar angle $\phi$), provide an even more useful basis. The helicity operator commutes with the Dirac Hamiltonian, has the straightforward physical interpretation of normalised spin along particle flight direction, and in the ultra-relativistic (i.e. massless) limit it is equivalent to chirality.

\bibliography{main}

\end{document}